\documentclass{optica-article}

\journal{opticajournal} 

\articletype{Research Article}

\usepackage{lineno}
\usepackage{xcolor}
\usepackage{dsfont}
\hypersetup{colorlinks=true,citecolor={blue},linkcolor={blue},urlcolor={blue}}

\begin{document}

\definecolor{lilac}{RGB}{150,100,180}
\newcommand{\Stella}[1]{{\color{lilac}{#1}}}
\newcommand{\hs}[1]{\textcolor{purple}{#1}}

\title{Polariton vortex Chern insulator}

\author{S. L. Harrison,\authormark{1} A. Nalitov,\authormark{2,3,4} P. G. Lagoudakis,\authormark{5,1} and H. Sigurðsson\authormark{6,*}}

\address{\authormark{1}School of Physics and  Astronomy, University of Southampton, Southampton, SO17 1BJ, United Kingdom\\
\authormark{2}Faculty of Science and Engineering, University of Wolverhampton, Wulfruna Street, Wolverhampton WV1 1LY, United Kingdom\\
\authormark{3}Moscow Institute of Physics and Technology, Institutskiy per., 9, Dolgoprudnyi, Moscow Region, Russia 141701\\
\authormark{4}
ITMO university, Kronverksky pr., 49, lit. A, St. Petersburg, Russian Federation\\
\authormark{5}Hybrid Photonics Laboratory, Skolkovo Institute of Science and Technology, Territory of Innovation Center Skolkovo, 6 Bolshoy Boulevard 30, Bld. 1, 121205 Moscow, Russia\\
\authormark{6}Science Institute, University of Iceland, Dunhagi 3, IS-107, Reykjavik, Iceland
}

\email{\authormark{*}helg@hi.is} 


\begin{abstract*} 
We propose a vortex Chern insulator, motivated by recent experimental demonstrations on programmable arrangements of cavity polariton vortices by Alyatkin et al.~\cite{alyatkin_all-optical_2022} and Wang et al.~\cite{Wang_NSR2022}. In the absence of any external fields, time-reversal symmetry is spontaneously broken through polariton condensation into structured arrangements of localized co-rotating vortices. We characterize the response of the rotating condensate lattice by calculating the spectrum of Bogoliubov elementary excitations and observe the crossing of edge-states, of opposite vorticity, connecting bands with opposite Chern numbers. The emergent topologically nontrivial energy gap stems from inherent vortex anisotropic polariton-polariton interactions and does not require any spin-orbit coupling, external magnetic fields, or elliptically polarized pump fields.
\end{abstract*}

\section{Introduction}
Topological Chern insulators are exotic materials with a bulk band gap but gapless edge states that experience topologically robust protection from scattering due to time reversal symmetry~\cite{hasan_colloquium_2010,bansil_colloquium_2016,qi_topological_2011,chiu_classification_2016}. The difference between an ordinary- and a topological insulator (TI) is manifested in the famous quantum Hall effect describing quantization of edge conductance in the presence of a magnetic field~\cite{thouless_quantized_1982}. The origin of which comes from twisting Berry phase of the Bloch bands that accumulates, over the Brillouin zone, in an integer topological invariant called the \textit{Chern number}. Nearly two decades ago, the quantum spin Hall effect was demonstrated in HgTe quantum wells~\cite{bernevig_quantum_2006, konig_quantum_2007}. Since then, topological properties have been explored in other electronic structures~\cite{kane_quantum_2005,fu_topological_2007,moore_topological_2007,xia_observation_2009,zhang_topological_2009}, ultracold bosonic~\cite{Aidelsburger_NatPhys2015} and fermionic~\cite{Jotzu_Nature2014} gases, photonic systems \cite{haldane_possible_2008,wang_reflection-free_2008,hafezi_robust_2011,raghu_analogs_2008,fang_microscopic_2011,wang_observation_2009,poo_experimental_2011,khanikaev_photonic_2013} and more recently in microcavity exciton-polariton (from here on \textit{polariton}) systems \cite{klembt_exciton-polariton_2018,bleu_interacting_2016, gulevich_exploring_2017, kartashov_bistable_2017, bleu_photonic_2017, mandal_antichiral_2019,st-jean_lasing_2017, nalitov_polariton_2015, sigurdsson_spontaneous_2019}.

Polaritons are hybrid light-matter quasiparticles appearing in the strong coupling regime of planar microcavities~\cite{Carusotto_RMP2013}. They possess extremely light effective mass and strong interactions making them highly suitable to explore out-of-equilibrium Bose-Einstein condensation at the interface of photonics and condensed matter, and elevated temperatures~\cite{Kasprzak_Nature2006}. Quantum fluids are known to be a source of fascinating dynamical behaviour~\cite{Leggett_Science2008} with polariton condensates being no exception when combined with TIs~\cite{bleu_interacting_2016, sigurdsson_spontaneous_2019}. Previous works have proposed that TIs can spontaneously emerge in the elementary excitation spectrum (also know as Bogoliubov excitations) in structured polariton fluids containing nontrivial hopping phases~\cite{bardyn_chiral_2016, Sigurdsson_PRB2017} in analogy to the Haldane model~\cite{haldane_model_1988}. Motivated by these past developments and the recent experimental demonstrations of optically~\cite{alyatkin_all-optical_2022} and lithographically patterned polariton vortex lattices~\cite{Wang_NSR2022} (as well vortex arrays in solid state lasers~\cite{Piccardo_NatPhot2022} and plasmon polariton systems~\cite{Kuo_JourOpt2018}) we develop a tight binding theory for localized polariton vortices. We apply our model to the honeycomb lattice and demonstrate a \textit{vortex Chern insulator} and outline some close analogies to the conventional spinor polariton Chern insulator~\cite{nalitov_polariton_2015, klembt_exciton-polariton_2018}. Our findings underpin a nonlinear optical system (i.e., a quantum fluid of polaritons) hosting both chiral and topologically protected edges modes which offers perspectives on optical-based nontrivial signal processing schemes and uni-directional flow of information.

A cornerstone of our work is the unique spatial coupling mechanism between two optically trapped polariton condensates populating the $l = \pm1$ orbital angular momentum (OAM) modes of each trap which was investigated by Cherotchenko et al. \cite{cherotchenko_optically_2021}. It was shown that ballistic expansion and overlapping of adjacent trapped condensates results in a dynamic dissipative coupling mechanism which depended on the relative angle between traps. Moreover, the coupling between adjacent co-rotating vortices and anti-rotating vortices could be optically engineered to be of similar strength. This is in contrast to spinor polariton fluids in which opposite-spin coupling defined by splitting of transverse electric and transverse magnetic (TE-TM) cavity photon modes is much weaker than same-spin coupling, severely limiting the size of the topological gap opening~\cite{klembt_exciton-polariton_2018}.

\section{Results}
\subsection{Model of coupled polariton condensate vortices}
Exciton-polariton condensates are conventionally described in the mean field picture using the generalized Gross-Pitaevskii equation coupled to an equation describing the dynamics of an exciton reservoir~\cite{Carusotto_RMP2013}. Instead of solving a 2+1 dimensional nonlinear partial differential equation, we project the condensate order parameter onto a truncated basis of the first-excited angular harmonics $l = \pm 1$ localized within each trap in the lattice. This gives $2 N_l$ coupled ordinary differential equations where $N_l$ is the number of lattice sites [see Supplementary Material for derivation],
\begin{equation}
i \frac{d\psi_{n,\pm}}{d t}  = \left[i p + (\alpha - i ) (|\psi_{n,\pm}|^2 + 2 |\psi_{n,\mp}|^2)   \right] \psi_{n,\pm} - \sum_{\langle n,m \rangle} \left[J \psi_{m,\pm} + \mathcal{J} \psi_{m,\mp} e^{\mp 2i \Theta_{n,m}}\right].
\label{eq.latt}
\end{equation}
Here, $\psi_{n,\pm}$ is the phase and amplitude of the $n$th condensate component with OAM $l=\pm 1$, $J$ and $\mathcal{J}$ are the "tunneling" rates between co-rotating and counter-rotating vortices, $\alpha$ corresponds to the repulsive polariton-polariton interactions, the negative imaginary term represents a gain saturation mechanism in the adiabatic exciton-reservoir limit~\cite{Carusotto_RMP2013}, and $p$ is the combined non-resonant optical pumping rate and cavity losses. From here on we scale time and other parameters in units of $J^{-1}$. 

\begin{figure}
    \centering
    \includegraphics[width=8.4cm]{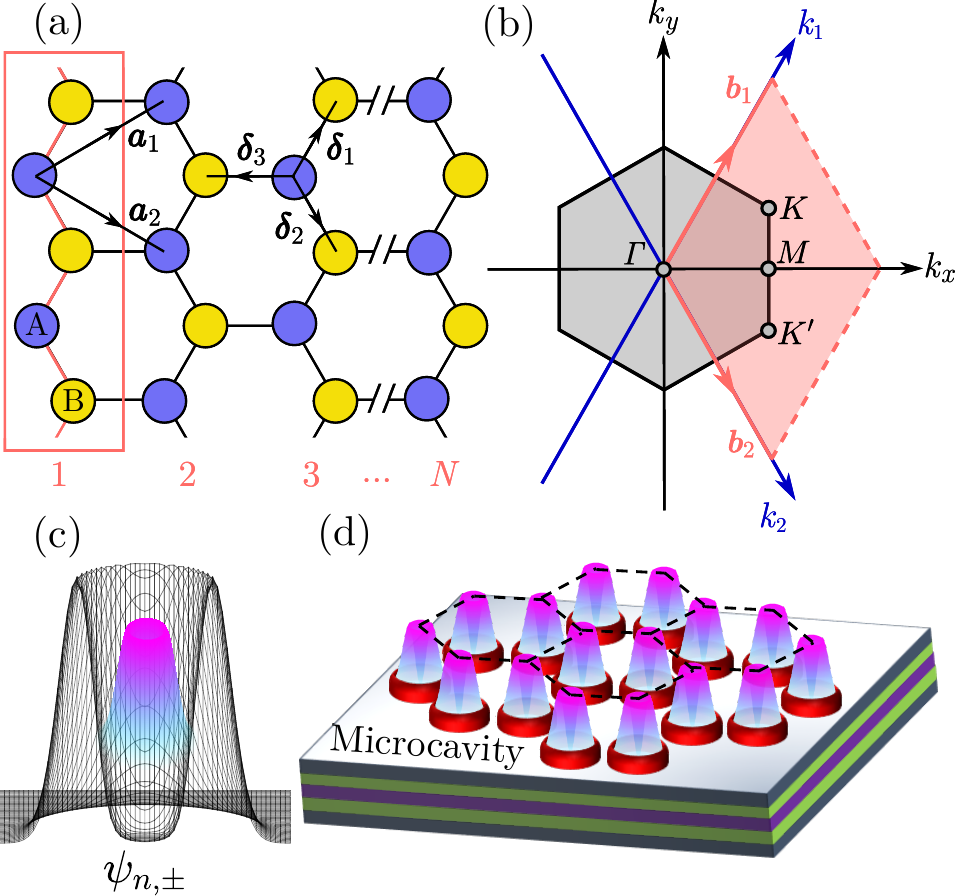}
    \caption{
     (a) Stripe honeycomb lattice consisting of two intercalated triangular lattices with $A$ and $B$ lattices sites shown in blue and yellow respectively. $\boldsymbol{a}_1$ and $\boldsymbol{a}_2$ are the lattice unit vectors, and $\boldsymbol{\delta}_i$, $i = \left\{1,2,3\right\}$ are the nearest-neighbour translations. A break in the lattice between chain 3 and $N$ is represented by $//$, where $N$ is even in this schematic. (b) Momentum space of the bulk lattice with both Cartesian and reciprocal lattice coordinate systems indicated. The reciprocal lattice vectors are shown by $\boldsymbol{b}_{1,2}$ in red, mapping out a diamond-shaped Brillouin zone, in addition to the hexagonal Brillouin zone shown by the grey hexagon. (c) Schematic showing the profile of the trapped vortex or (antivortex) $|\psi_{\pm}|^2$ (coloured profile). (d) Schematic of a polariton condensate vortex lattice injected and contained within ring-shaped optical traps (red) in a honeycomb structure (black dashed lines).}
    \label{fig:lattice}
\end{figure}

The sum runs over nearest neighbours and $\Theta_{n,m}$ is the angle of the link between two condensates in separate traps~\cite{cherotchenko_optically_2021} [i.e., angle of the edges connecting lattice sites in Fig.~\ref{fig:lattice}a]. The double-angle dependence of the coupling $\mathcal{J}$ between spatially separate counter-rotating vortices is analogous to a photonic spin-orbit coupling (SOC) mechanism in patterned cavities with TE-TM splitting~\cite{Sala_PRX2015}. In fact, this turns out to be one of the needed ingredients to obtain topological gap opening in spinor polariton Chern insulators~\cite{nalitov_polariton_2015,bardyn_topological_2015,klembt_exciton-polariton_2018,sigurdsson_spontaneous_2019}. The directionally dependent coupling in~\eqref{eq.latt} not only offers perspectives on using vortices to simulate spinor polariton TIs, but also spin transport phenomena under strong SOC ($J<\mathcal{J}$) in artificial polariton lattices~\cite{HeeBong_AQT2022}.

To observe the emergence of topologically protected edge states in the Bogoliubov elementary excitations of the polariton vortex fluid we will work with a honeycomb lattice geometry, $\Theta_{n,m} \in \{\pi, \pm \pi/3\}$, hosting Dirac points. To open the gap at those points, time reversal symmetry must be broken~\cite{hasan_colloquium_2010}. Typically, the gap can be opened for honeycomb polaritons through direct methods such as applying a normal-incident magnetic field or an elliptically polarized excitation source resulting in either real or effective Zeeman fine-structure splitting of spinor polaritons~\cite{nalitov_polariton_2015,klembt_exciton-polariton_2018}. However, for truly emergent TIs, the breaking of time-reversal symmetry is spontaneous corresponding to a many-body effect~\cite{Sigurdsson_PRB2017, sigurdsson_spontaneous_2019} such as condensation into a macroscopic coherent state which can possess nontrivial chiral phase and density profile~\cite{alyatkin_all-optical_2022}.

\subsection{The Honeycomb Lattice}
The honeycomb lattice is made up of a two-site unit cell. Each site in the unit cell is labelled $A$ and $B$ (shown by blue and yellow spots in \autoref{fig:lattice}), which when expanded following lattice vectors $\boldsymbol{a}_1$ and $\boldsymbol{a}_2$ makes a structure of many tessellating hexagons with a nearest neighbour separation distance of $a$. Here, the real space lattice vectors are defined as:
\begin{equation}
    \boldsymbol{a}_1 = \frac{a}{2} (3,\sqrt{3}),~~~~\boldsymbol{a}_2 = \frac{a}{2}(3,-\sqrt{3}),
    \label{eq:a1_a2}
\end{equation}
%
and the reciprocal lattice vectors as:
\begin{equation}
    \boldsymbol{b}_1 = \frac{2\pi}{3a}(1,\sqrt{3}), \qquad \boldsymbol{b}_2 = \frac{2\pi}{3a}(1,-\sqrt{3}).
\end{equation}
The hexagonal structure of the honeycomb lattice leads to a hexagonal structure in its Brillouin zone. The Dirac points are located at $\boldsymbol{K}$ and $\boldsymbol{K}^\prime$, as shown in \autoref{fig:lattice}, whose positions in momentum space are:
\begin{equation}
    \boldsymbol{K} = \bigg(\frac{2\pi}{3a},\frac{2\pi}{3\sqrt{3}a}\bigg),~~~\boldsymbol{K}^\prime = \bigg(\frac{2\pi}{3a},-\frac{2\pi}{3\sqrt{3}a}\bigg).
\end{equation}
The lattice strip consists of $N$ zigzag chains (see box in Fig.~\ref{fig:lattice}b) along the $x$-axis and infinite along the $y$ axis. The index of each chain is labelled by a red number.

\subsection{"Ferromagnetic" vortex solution}
In order to spontaneously break time reversal symmetry in the honeycomb lattice the condensate needs to form into a solution with net OAM. There exist two particularly simple fixed point solutions to~\eqref{eq.latt} under periodic boundary conditions written,
\begin{equation} \label{eq.sol0}
\psi_{n_A,l}  = \pm \psi_{n_B,l} = \sqrt{p}, \qquad \psi_{n,-l} = 0, 
\end{equation} 
Here, $n_A$ and $n_B$ refer to sites on sublattice $A,B$. These two solutions are referred to as the FM solutions since every site in the lattice is equally populating only the $l$ vortex component while the other component is zero. Physically, each site has a condensate with equal OAM pointing in the same direction with all nearest neighbours either in-phase ($\psi_{n_A,l}  = \psi_{n_B,l}$) or anti-phase ($\psi_{n_A,l}  = -\psi_{n_B,l}$). 


\subsection{Bogoliubov elementary excitations ({\it bogolons})}
\subsubsection{Linearization of the condensate equations of motion}
In this section we linearize the coupled condensate equations of motion~\eqref{eq.latt} and derive an eigenvalue problem for the spectrum of the Bogoliubov elementary excitations (also referred to as {\it bogolons}). We then analyse the bogolon spectrum, around the aforementioned FM solutions, an look for evidence of topologically protected edge states. The perturbed condensate solution, in momentum space, can be written
\begin{equation}
\psi'_{\mu,l}(\boldsymbol{k}) = [\psi_{\mu,l}(\boldsymbol{k}) + \delta\psi_{\mu,l}(\boldsymbol{k})] e^{-i \nu t}, \qquad \qquad \mu = A,B
\end{equation}
where $\hbar \nu$ is the energy of the condensate. The momentum space representation is obtained through the standard Fourier transform of the order parameters $\psi_{n_\mu,l} \propto \sum_{\mathbf{k}} e^{i \mathbf{k} \cdot \mathbf{r}_{n_\mu}} \psi_{\mu,l}(\mathbf{k})$. The perturbation is expanded in the general fashion,
\begin{equation} \label{eq.fluc}
 \delta\psi_{\mu,l}(\boldsymbol{k}) = u_{\mu,l} e^{i (\boldsymbol{k} \cdot \boldsymbol{r} - \omega t)} + v_{\mu,l}^* e^{-i (\boldsymbol{k}\cdot \boldsymbol{r} - \omega^* t)}.
\end{equation}
%
Plugging~\eqref{eq.fluc} into the condensate equations of motion~\eqref{eq.latt} and keeping only terms linear in $\delta\psi_{l}$ results in an $8\times8$ eigenvalue problem $\omega \boldsymbol{\delta \psi}=\mathbf{M}\boldsymbol{\delta \psi}$ in momentum space in the basis
$\boldsymbol{\delta \psi}=[u_{A,+},v_{A,+},u_{A,-},v_{A,-},u_{B,+},v_{B,+},u_{B,-},v_{B,-},]^\mathrm{T}$. The matrix $\mathbf{M}$ describes the evolution of the perturbation (analogous to the Jacobian for coupled ordinary differential equations) and can be written neatly in a $2\times2$ block form,
\begin{equation} 
\mathbf{M} = \begin{pmatrix} \mathcal{M}_A & \mathbf{J}_{\boldsymbol{k}}  \\
\mathbf{J}_{\boldsymbol{k}}^\dagger   &  \mathcal{M}_B 
   \end{pmatrix}.
   \label{eq:Minfinite}
\end{equation}
The diagonal blocks depend on the condensate solution and are decomposed into the following parts,
\begin{equation}
\mathcal{M}_{\mu} =   ip \mathds{I}_{4\times4} + 2\alpha\mathcal{M}_{\mu,\alpha} -  2i  \mathcal{M}_{\mu,R}.
\end{equation}
The first term describes the gain of the system. The second and third terms describe the real and imaginary contribution to the bogolon energies from the nonlinear terms in~\eqref{eq.latt}, respectively: 
\begin{equation} \label{eq.alpha}
        \mathcal{M}_{\mu,\alpha} =
        \begin{pmatrix} 
        |\psi_{\mu,+}|^2 + |\psi_{\mu,-}|^2  &  \frac{1}{2}(\psi_{\mu,+})^2 & \psi_{\mu,+}\psi_{\mu,-}^* & \psi_{\mu,+}\psi_{\mu,-} \\
		- \frac{1}{2}(\psi_{\mu,+}^*)^2 & - |\psi_{\mu,+}|^2 - |\psi_{\mu,-}|^2  & -\psi_{\mu,+}^*\psi_{\mu,-}^* & -\psi_{\mu,+}^*\psi_{\mu,-} \\
		\psi_{\mu,+}^*\psi_{\mu,-} & \psi_{\mu,+}\psi_{\mu,-} &  |\psi_{\mu,+}|^2 + |\psi_{\mu,-}|^2&  \frac{1}{2}(\psi_{\mu,-})^2 \\
		-\psi_{\mu,+}^*\psi_{\mu,-}^* &  -\psi_{\mu,+}\psi_{\mu,-}^* & -  \frac{1}{2}(\psi_{\mu,-}^*)^2 & - |\psi_{\mu,+}|^2 -  |\psi_{\mu,-}|^2 
		\end{pmatrix},
\end{equation}
\begin{equation} \label{eq.R}
    \mathcal{M}_{\mu,R} =
    \begin{pmatrix} 
        1&0&0&0 \\
        0&-1&0&0 \\
        0&0&1&0 \\
        0&0&0&-1 
		\end{pmatrix} \mathcal{M}_{\mu,\alpha}.
\end{equation}
The off-diagonal blocks $\mathbf{J}_{\boldsymbol{k}}$ describe coupling between the sublattices [see Supplementary Material],
\begin{equation}
\mathbf{J}_{\boldsymbol{k}} = (\hat{J}_{0} + \hat{J}_{1} e^{-ik_1} + \hat{J}_{2} e^{-ik_2}) \otimes \hat{\sigma}_z
\label{eq:Jkri_infinite}
\end{equation}
where $k_{1,2}$ span the Brillouin zone along $b_{1,2}$ respectively, [see \autoref{fig:lattice}(b)], and

\begin{equation}
\hat{J}_{n} = 
-\begin{pmatrix}
J & \mathcal{J} e^{-2i\Theta_n}  \\
\mathcal{J} e^{2i\Theta_n} & J
\end{pmatrix}, \qquad n = 0,1,2
\end{equation}
%
%
Here, $\Theta_n$ is the momentum space equivalent of $\Theta_{n,m}$ in~\eqref{eq.latt}. The angles between nearest neighbours are $\Theta_0 = \pi$, $\Theta_1 = \pi/3$ and $\Theta_2 = -\pi/3$.

\subsubsection{PT symmetry phase transition}
One feature of the bogolon operator~\eqref{eq:Minfinite} is its inherent non-Hermitian structure as can be seen from just the matrix~\eqref{eq.alpha}. Non-Hermitian topological physics has seen a lot of interest in the recent years~\cite{Ding_NatRevPhys2022} with enriched classification of topological phases of matter~\cite{Yao_PRL2018}, exceptional points~\cite{Bergholtz_RMP2021}, and the non-Hermitian skin effect~\cite{Xu_PRB2021}. However, for the purpose of this study, we will consider a FM vortex lattice in which the parameters are chosen such that the bogolon dispersion is purely real-valued up to a constant imaginary factor and with separable bulk bands [i.e., $\omega_n(\mathbf{k}) \neq \omega_m(\mathbf{k})]$ for all bands $m\neq n$). This then precludes the need for specialized treatment to classify non-Hermitian topological phases in the considered vortex lattice~\cite{Bergholtz_RMP2021} which will become a subject of a later study.
\begin{figure}[t]
    \centering
    \includegraphics[width=7.5cm]{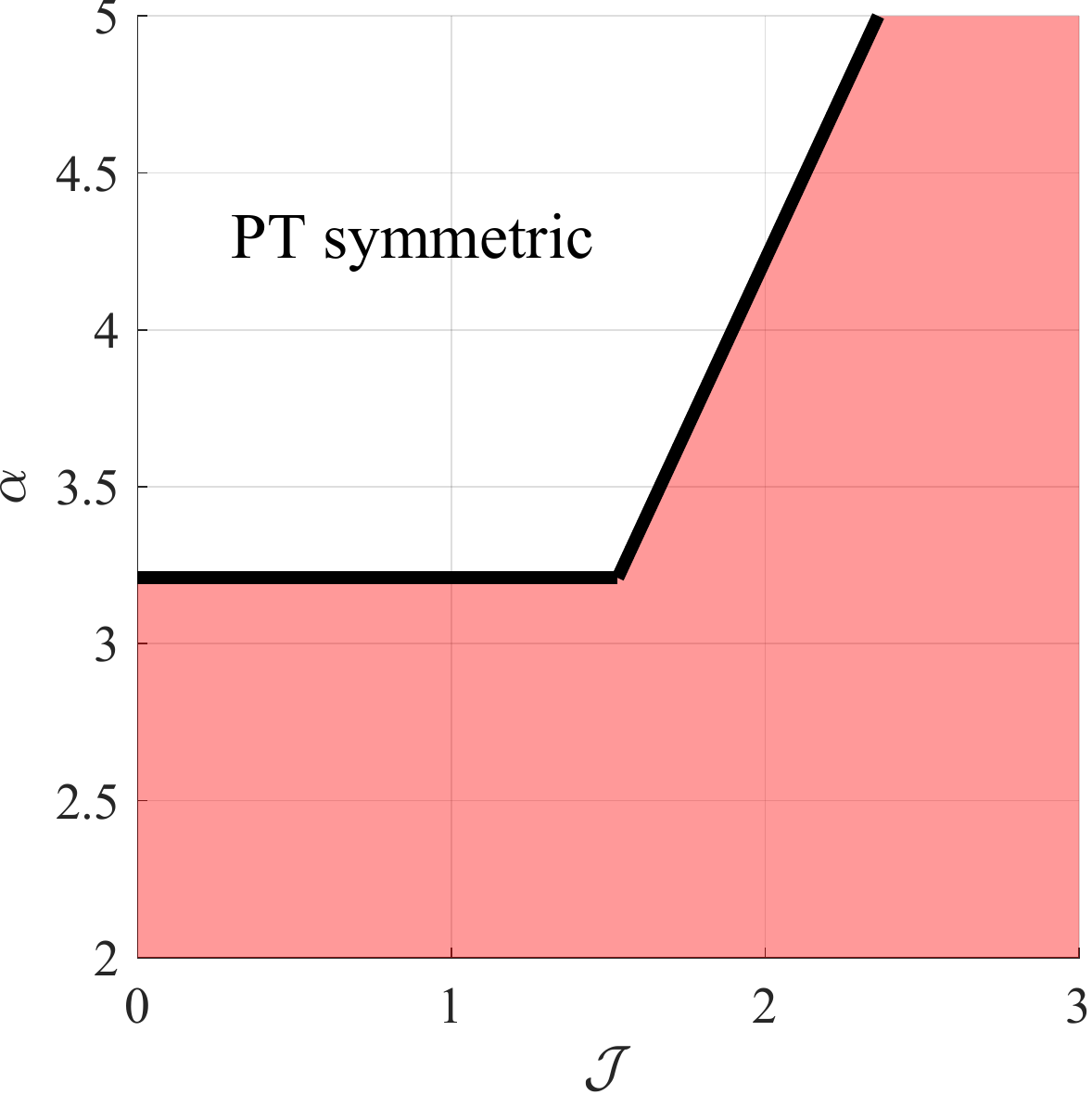}
    \caption{Bogolon PT symmetry boundary for the FM vortex solution for $p=1$.}
    \label{PT_phase_map}
\end{figure}

It is worth noting that non-Hermitian linear operators with a completely real spectrum are referred to as parity-time (PT) symmetric operators. Such operators have also gathered a lot of attention recently, especially in photonic systems with tailored gain and loss landscapes~\cite{Ozdemir_NatMat2019}. Figure~\ref{PT_phase_map} shows the boundary of the PT symmetric phase of the FM vortex lattice in the $\alpha$-$\mathcal{J}$ plane obtained by diagonalizing~\eqref{eq:Minfinite}. The results presented within the current study work within this boundary.

\subsubsection{Bogolon dispersion relation}
\begin{figure}
    \centering
    \includegraphics[width=8.4cm]{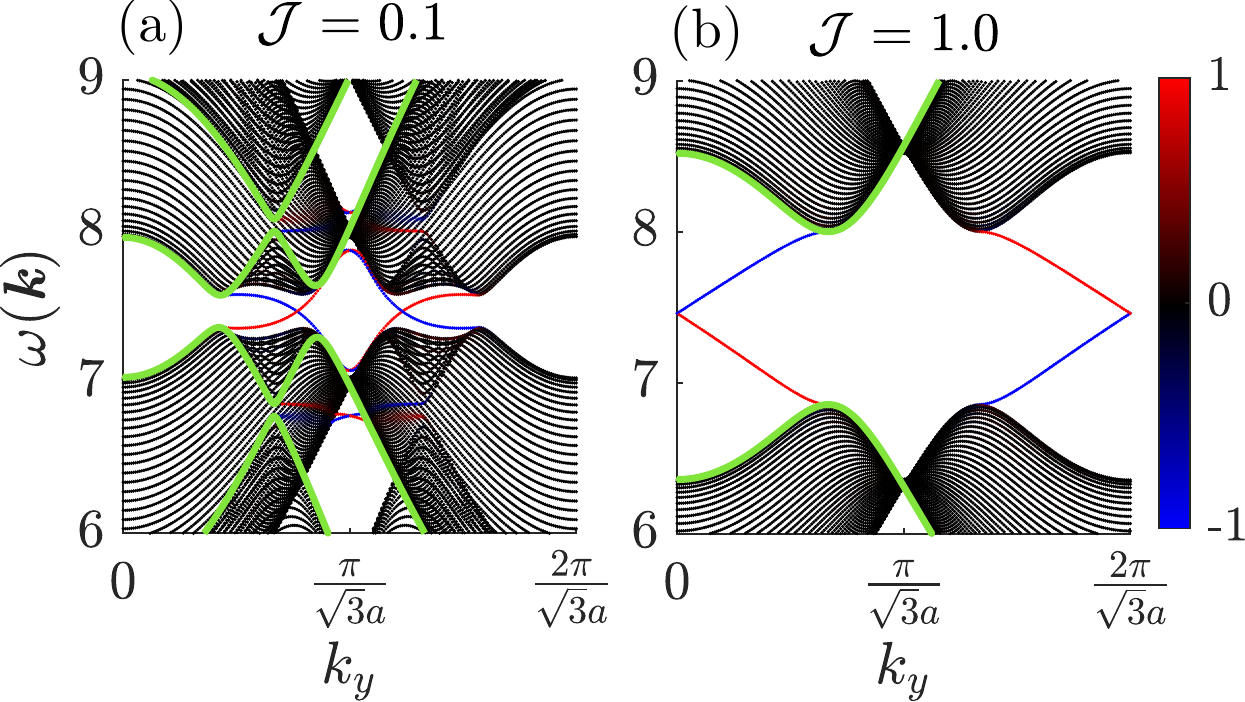}
    \caption{
    Positive energy bogolon dispersion relation obtained from diagonalizing~\eqref{eq:Minfinite} for both the (green line) infinite and (blue/black/red) $N=50$ finite honeycomb strip. The blue-red colours represent localisation of the eigenstates on the left and right lattice edges, respectively. In all plots, $k_x = 2\pi/3a$, $\alpha = 4$, and $p = 1$.
    }
    \label{fig:SpinVsBog}
\end{figure}

Plugging the FM solution~\eqref{eq.sol0} into our Bogoliubov operators~\eqref{eq.alpha} and~\eqref{eq.R} we proceed to diagonalize~\eqref{eq:Minfinite} for a $N=50$ honeycomb lattice strip (periodic boundaries along the $y$ axis). The bogolon dispersion relation for two different values of $\mathcal{J}$ is shown in \autoref{fig:SpinVsBog} where we observe four and two chiral edge states, respectively. The blue and red colors indicate localization of the bogolon on the left and right edges respectively. The transition from four to two edge states is a topological transition in which the Chern numbers of the upper and lower bands change from $C_n=\pm2$ to $C_n = \pm1$ as we verify in the next section. Physically, in the absence of the condensate $\psi_{n,l}=0$ additional Dirac points appear at the $M$-point of the lattice for a critical value of $\mathcal{J}=1/2$ causing a change in the dispersion topology~\cite{nalitov_polariton_2015}.

To verify that the gap is complete in the bulk we diagonalize~\eqref{eq:Minfinite} for periodic boundaries along both $x$ and $y$. The corresponding "infinite lattice" bulk dispersion is shown with green solid lines which indeed do not cross. The blue and red colors indicate localization of the bogolon on the left and right edges respectively. For completeness, in the Supplementary Material we show that our results qualitatively, and nearly quantitatively, match the conventional spinor polariton TI which utilizes TE-TM splitting and a real external magnetic field~\cite{nalitov_polariton_2015, bardyn_topological_2015 ,klembt_exciton-polariton_2018}. 

The presence of chiral edge states in the dispersion implies that bogolons experience the FM vortex lattice as an effective out-of-plane magnetic field that breaks time-reversal symmetry with consequent nontrivial topological gap opening. This effect can be understood intuitively by scrutinizing the energy shift coming from the condensate onto the bogolons. For the case of $l=1$ FM vortex lattice, equation~\eqref{eq.alpha} in the half-basis $[u_+, v_+, u_-, v_-]^\text{T}$ becomes simply
\begin{equation} \label{eq.alpha2}
         \mathcal{M}_{\alpha} =
        p\begin{pmatrix} 
        1 & 1/2 & 0 & 0 \\
        -1/2 & -1 & 0 & 0 \\
        0 & 0 & 1 & 0 \\
        0 & 0 & 0 & -1
		\end{pmatrix}.
\end{equation}
The off-diagonal terms mix together $l=1$ positive and negative energy bogolons ($|u_+\rangle$ and $|v_+\rangle$) while $l=-1$ bogolons ($|u_-\rangle$ and $|v_-\rangle$) are still good eigenstates. We have dropped the sublattice index $A,B$ for now since both sublattice matrices are identical. The eigenvalues of~\eqref{eq.alpha2} are $\omega_{1,2} = \pm \sqrt{3}/2$ and $\omega_{3,4} = \pm 1$ and the corresponding eigenvectors,
\begin{equation}
    \mathbf{v}_{1} = \begin{pmatrix}
    2+\sqrt{3} \\
    -1 \\
    0 \\
    0
    \end{pmatrix}, \qquad
    \mathbf{v}_{2} = \begin{pmatrix}
    -2+\sqrt{3} \\
    1 \\
    0 \\
    0
    \end{pmatrix}, \qquad \mathbf{v}_{3}  =  \begin{pmatrix}
    0 \\
    0 \\
    1 \\
    0
    \end{pmatrix}, \qquad \mathbf{v}_{4}  =  \begin{pmatrix}
    0 \\
    0 \\
    0 \\
    1
    \end{pmatrix}.
\end{equation}
Therefore, the background vortex lattice splits the energies of $l = \pm 1$ bogolons by amount $\Delta = \omega_{3,4} - \omega_{1,2}$ similar to a spin in a magnetic field. When the condensate rotation within the lattice reverses the splitting also reverses. As opposite vortices couple across the sublattices ($\mathcal{J}\neq0$) the band degeneracy is removed and a topologically nontrivial gap opens around the Dirac points.
%



\subsubsection{Chern numbers}
To further verify that the edge states observed in Fig.~\ref{fig:SpinVsBog} are topologically nontrivial in nature we calculate the Berry curvature and the associated Chern numbers of the upper and lower bulk bogolon bands. For band $n$, with normalised eigenvector $\vert \phi_{n} (\boldsymbol{k}) \rangle$, the Berry curvature in two dimensions is, 
 \begin{equation}
 \mathbf{B}_n(\boldsymbol{k}) = \frac{\partial A_{n,y}(\boldsymbol{k})}{\partial k_x} - \frac{\partial A_{n,x}(\boldsymbol{k})}{\partial k_y},\label{Eq:berryCurve}
 \end{equation}
 where $A_{n,x(y)}(\boldsymbol{k})$ is the Berry connection along the $x(y)$ direction,
 \begin{equation}
 A_{n,x(y)}(\boldsymbol{k}) = i\langle \phi_{n}(\boldsymbol{k})\vert\partial_{k_{x(y)}}\vert \phi_{n}
 (\boldsymbol{k}) \rangle.
\label{Eq:berryConnection}
 \end{equation}
The Chern number of the $n$th band can be found from integrating over the Berry curvature within the first Brillouin zone,
 \begin{equation} \label{eq.chern}
     C_n = \frac{1}{2\pi i}\iint_\text{BZ}\mathbf{B}_n(\boldsymbol{k}) dk_x dk_y
 \end{equation}
In order to calculate the Berry curvature and the associated Chern number of the lattice bands on a discrete $\mathbf{k}$-space grid we use the approach of Fukui et al. \cite{fukui_chern_2005}.

\begin{figure}
    \centering
    \includegraphics[width=8.4cm]{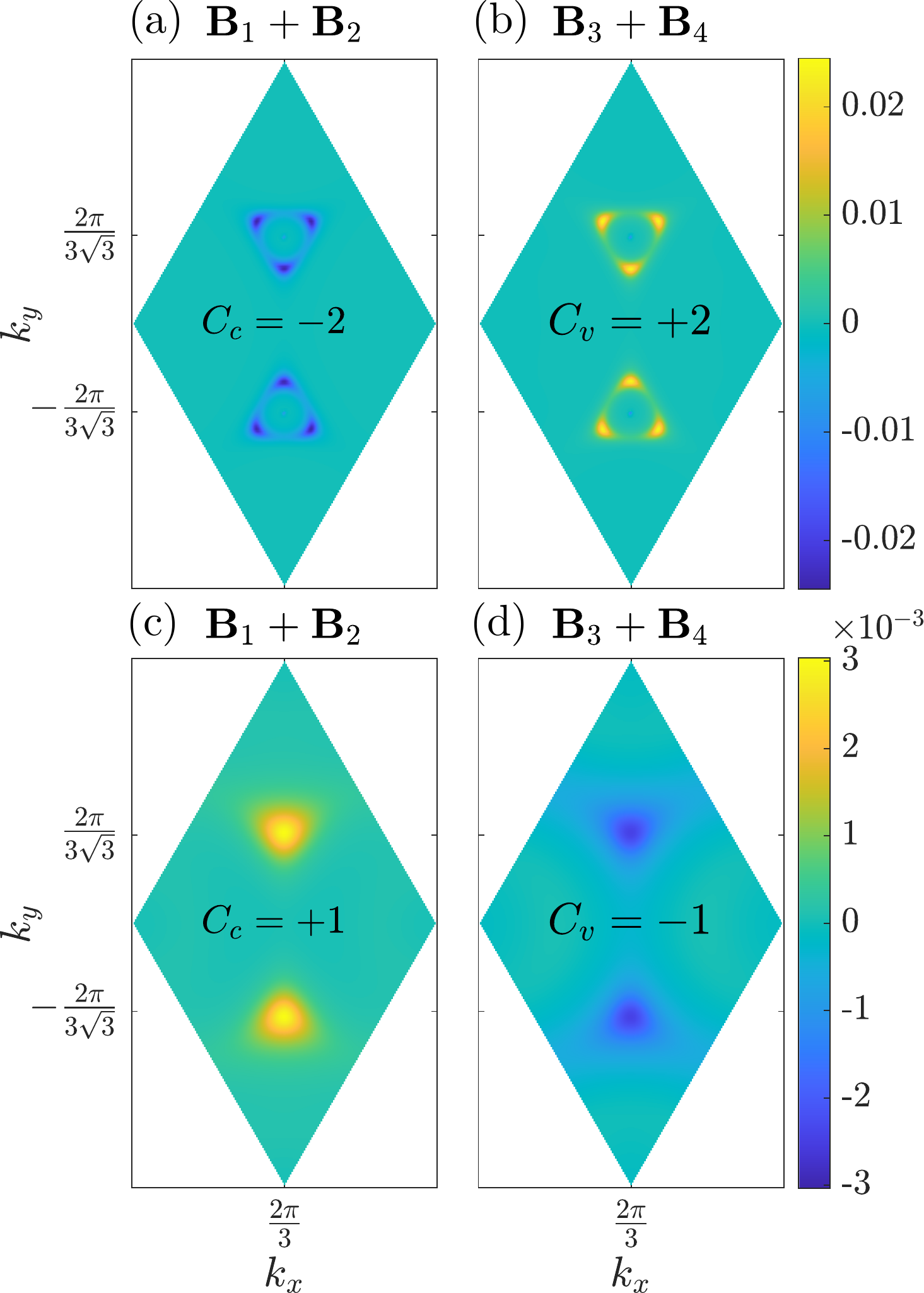}
    \caption{Berry curvature summed over (counting from the highest energy band) (a,c) the first pair and (b,d) the second pair of bands of the infinite honeycomb vortex lattice following from Bogoliubov elementary excitation with (a,b) $\mathcal{J} = 0.1$ and (c,d) $\mathcal{J} = 1.0$. The Brillouin zone follows the red diamond in \autoref{fig:lattice}(c). Here, $\alpha = 4$ and $p = 1$.}
    \label{fig:berryBog}
\end{figure}

We show results only for the positive energy branches of the bogolon dispersion since the negative branches (also referred to as \textit{ghost branches}) are just reflected about $\omega=0$. Counting the bands from top to bottom, the cumulative Berry curvature and Chern numbers for the positive energy ``conduction'' and ``valence'' bands are written $\mathbf{B}_{c(v)} =  \mathbf{B}_{1(3)} + \mathbf{B}_{2(4)}$ and $C_{c(v)} = C_{1(3)} + C_{2(4)}$ respectively. The results are shown in \autoref{fig:berryBog} for $\mathcal{J}=0.1$ and $\mathcal{J}=1.0$. For weaker opposite-vortex coupling strengths a clear trigonal warping in the Berry curvature, localised around the $\boldsymbol{K}$ and $\boldsymbol{K}^\prime$ points, can be observed which has been reported before for spinor polariton TIs~\cite{bleu_interacting_2016}. From~\eqref{eq.chern} we obtain band Chern numbers of $C_c = -2$ and $C_v = +2$ which, when the bulk-boundary correspondence is invoked~\cite{hasan_colloquium_2010}, results in four topologically protected edge states crossing the gap as seen in Fig.~\ref{fig:SpinVsBog}(a). At higher opposite-vortex coupling strengths the system undergoes a topological transition where additional Dirac cones form and annihilate in pairs, reducing the Chern number from $\pm2$ to $\pm1$. Our calculations of the band Chern numbers verify that the bogolon gap opening is an emergent topologically nontrivial transition induced by the vortex lattice~\cite{bardyn_chiral_2016, Sigurdsson_PRB2017}.

\subsection{Numerical Simulations}
To demonstrate the chiral response of the bogolons at the edges of the FM vortex honeycomb strip we perform numerical simulations of~\eqref{eq.latt} under resonant injection. Here, we choose the $l=1$ antiphase FM solution in~\eqref{eq.sol0} as an initial condition for $\alpha=4$, $p=1$, and $\mathcal{J}=1$. Numerically solving the equations of motion~\eqref{eq.latt} in time we observe that the FM solution is indeed stable and only experiences slight density modulations at the edges of the $N=12$ honeycomb strip. 

\begin{figure}[t]
    \centering
    \includegraphics[width=8.4cm]{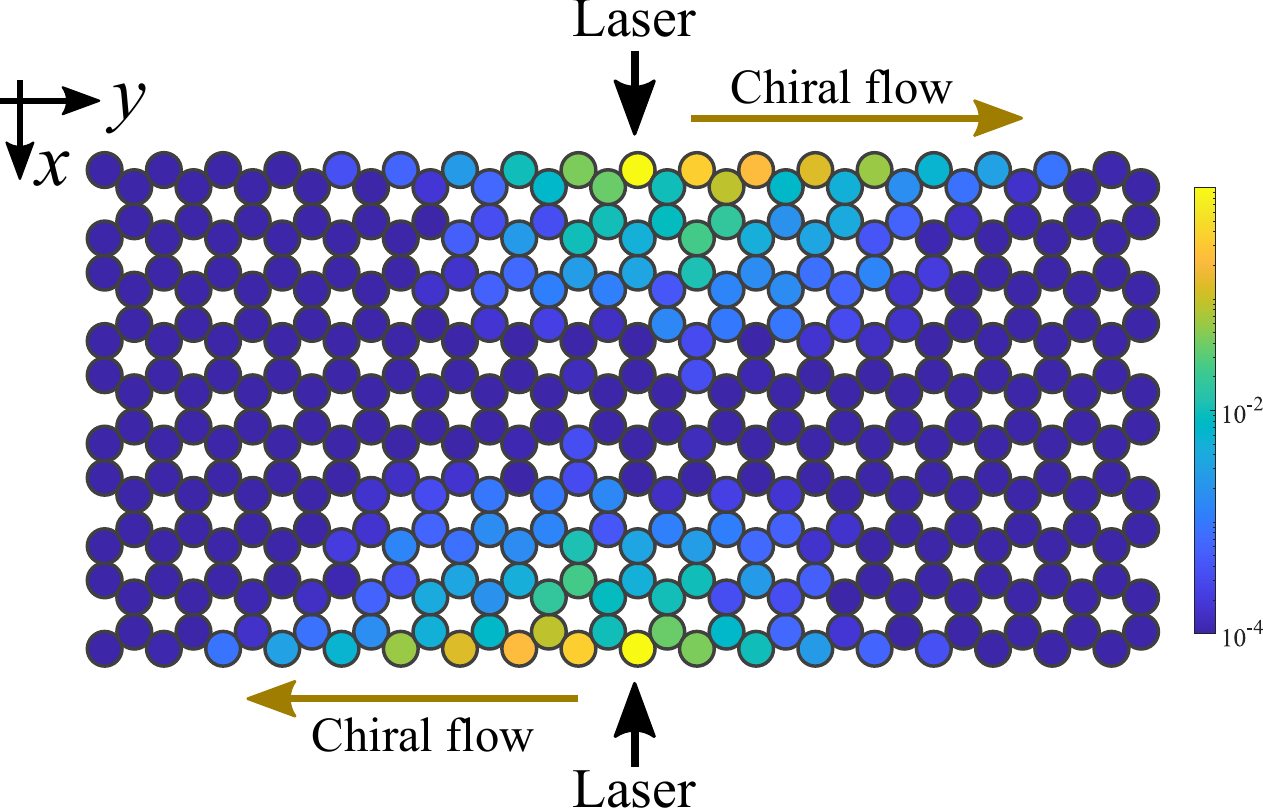}
    \caption{
    Simulation using~\eqref{eq.latt} showing a snapshot of the normalized antivortex component intensity $\vert \psi_{-}(t)\vert^2$ under a cw resonant injection at the indicated sites. The resulting density profile displays clear chiral response of the antivortex bogolons. Here the parameters are: $\alpha = 4$, $p = 1$, $\mathcal{J} = 1$, $P = 0.2$, and $\tilde{\omega} = 6.9$.
    }
    \label{fig:edgeModes}
\end{figure}

We then resonantly drive the antivortex bogolons ($l=-1$) at two lattice sites located at opposite edges of the strip (see black arrows in Fig.~\ref{fig:edgeModes}). The resonant excitation corresponds to a term $+P_n e^{-i \tilde{\omega} t}$ appended to the right hand side of~\eqref{eq.latt}. The resonant drive at the chosen lattice sites results in an antivortex perturbation which propagates in opposite directions at opposite edges due to the influence of the background $\psi_+$ component (not shown). We point out that the perturbations are attenuated, with a lifetime $\tau_\text{bog} = 1/p$, which is the reason they decay along the strip. 

\section{Conclusions}
We have designed a tight-binding model describing the coupling between localized polariton vortices and identify analogies with conventional spin-orbit-coupled spinor polaritons~\cite{Sala_PRX2015}. Our model offers a path towards simulating spinor polariton phenomena~\cite{HeeBong_AQT2022} with tunable nonlinearities and anistropic coupling strengths. In particular, we investigate the paradigmatic honeycomb lattice~\cite{nalitov_polariton_2015, klembt_exciton-polariton_2018}, possessing Dirac points, and show that when all vortices start rotating in the same direction time-reversal symmetry is spontaneously broken for the condensate Bogoliubov elementary excitations. A gap opens around the lattice Dirac points with the emergence of chiral topologically protected edge states bridging bulk bands of opposite Chern numbers. Our findings could play a role in optical based nonlinear topological signal processing schemes with uni-directional flow of vortical information~\cite{Lu_NatPho2014} and technologies based on symmetry protected phase-singular optics~\cite{Shen_LMAppl2019}.

 \section*{Funding}
	S.L.H. acknowledges the support of the UK's Engineering and Physical Sciences Research Council through a Doctoral Prize Award, Grant No.~EP/T517859/1. A.N. acknowledges support by the Russian Science Foundation under Grant No. 22-12-00144. H.S. and P.G.L. acknowledge the European Union's Horizon 2020 program, through a FET Open Research and Innovation Action under the Grant Agreement No.~899141 (PoLLoC) and No.~964770 (TopoLight). H.S. also acknowledges the Icelandic Research Fund (Rannis), Grant No.~239552-051.

 \section*{Disclosure}
 The authors declare no conflicts of interest.



\newpage
\setcounter{section}{0}
\setcounter{equation}{0}
\setcounter{figure}{0}
\setcounter{table}{0}
\renewcommand{\theequation}{S\arabic{equation}}
\renewcommand{\thefigure}{S\arabic{figure}}
\renewcommand{\thesection}{S\arabic{section}}

{\huge {\fontfamily{phv}\selectfont Supplemental document}}

\section{Derivation of the polariton vortex tight binding model}
The two dimensional generalized Gross-Pitaevskii equation (2DGPE) describing a polariton condensate wavefunction $\Psi(\mathbf{r},t)$ coupled to an exciton reservoir $n_X(\mathbf{r},t)$ is written~\cite{Carusotto_RMP2013},
\begin{align}
i  \hbar\frac{\partial \Psi}{\partial t} & = \left[ - \frac{\hbar^2 \nabla^2}{2m^*}  + \alpha |\psi|^{2} - \frac{i \hbar \gamma}{2} + \left( G + \frac{iR}{2} \right) n_X  + G\frac{P(\mathbf{r})}{W} + V(\mathbf{r}) \right] \Psi, \label{eq:GPE_psi}\\ 
\frac{\partial n_X}{\partial t} & = - (\Gamma + R |\Psi|^2) n_X + P(\mathbf{r}).
\label{eq:GPE_nx}
\end{align}
Here, $m^*$ is the polariton mass, $\gamma^{-1}$ the polariton lifetime, $G = 2 g |\chi|^2$ and $\alpha = g |\chi|^4$ are polariton-reservoir and polariton-polariton interaction strengths, respectively, $g$ is the exciton-exciton Coulomb interaction strength, $|\chi|^2$ is the excitonic Hopfield fraction of the polariton, $R$ is the scattering rate of reservoir excitons into the condensate, $\Gamma$ is the reservoir decay rate, $W$ quantifies additional blueshift coming from a background of high-energy and dark excitons generated by the nonresonant continuous-wave pump $P(\mathbf{r})$, and $V(\mathbf{r})$ represents the photonic potential landscapes in case of patterned cavities.

We will assume for simplicity that $V(\mathbf{r})=0$ and the reservoir steady state solution follows the condensate adiabatically so that $\partial_t n_X \simeq 0$ and expand the solution around small $R|\Psi|^2/\Gamma$,
\begin{equation}
n_X = \frac{P(\mathbf{r})}{\Gamma + R |\Psi|^2} \approx \frac{P(\mathbf{r})}{\Gamma} \left(1 - \frac{ R |\Psi|^2}{\Gamma} + \mathcal{O}{(|\Psi|^4)} \right).
\label{eq.X}
\end{equation}
This allows us to write a single partial differential equation for the condensate wavefunction,
\begin{equation}
i  \hbar \frac{\partial \Psi}{\partial t} = \left[ - \frac{\hbar^2 \nabla^2}{2m^*} + \alpha|\Psi|^2 - \frac{i \hbar \gamma}{2}  + \left(G + i \frac{R}{2}\right) \frac{P(\mathbf{r})}{\Gamma} \left(1 - \frac{R |\Psi|^2}{\Gamma} \right) + G \frac{P(\mathbf{r})}{W} \right] \Psi 
\label{eq.2DGPE}
\end{equation}
We are interested in an optically trapped condensate where the pump is radially symmetric, $P(\mathbf{r}) = P(r)$. It has been shown that polaritons in such optical traps can condense into the trap higher order modes~\cite{Askitopoulos_PRB2015} and specifically the degenerate pair of clockwise and anticlockwise orbital angular momentum (OAM) states~\cite{Ma_NatComm2020} which we write as,
\begin{equation}
\Psi(\mathbf{r},t) = \xi(r) \left( \psi_+ e^{i \varphi} + \psi_- e^{-i \varphi} \right) e^{-i E t/\hbar}.
\label{eq.basis}
\end{equation}
Here, $\xi(r)$ is the radial solution to $\partial_t |\Psi|^2 = 0$ for a condensate in a single optical trap. Plugging in this truncated solution and integrating over the space, exploiting the orthogonality of the states, we arrive at two coupled equation of motion describing the phase and occupation (amplitude) of each vortex component,
\begin{equation}
i \frac{d \psi_\pm}{d t}  = \left[i \tilde{p} + (\tilde{\alpha} - i \tilde{R}) (|\psi_\pm|^2 + 2 |\psi_\mp|^2)   \right] \psi_\pm,
\label{eq.cpm}
\end{equation}
up to an overall energy shift. The new coefficients are,
\begin{subequations}
\begin{align}	 \label{intp}
\tilde{p} & = \frac{R}{2 \hbar \Gamma} \int \xi(r)^2 P(r) \, d\mathbf{r} - \frac{ \gamma}{2}, \\
\tilde{\alpha} & = \frac{\alpha}{\hbar} \int \xi(r)^4 \, d\mathbf{r} -  \frac{gR}{\hbar\Gamma^2} \int \xi(r)^4 P(r) \, d\mathbf{r}, \\ \label{intR}
\tilde{R} & = \frac{R^2}{2 \hbar^2 \Gamma^2} \int \xi(r)^4 P(r) \, d\mathbf{r}.
\end{align}
\end{subequations}
In equation (1) in the main text we have dropped the tilde symbol for simplicity.

\subsection{Delta shell potential}
Useful identities can be derived by considering the pumping profile as a delta shell potential $P(r) = P_0 \delta (r-r_0)/r_0$. Collecting the terms in~\eqref{eq.2DGPE} proportional to the pump, we can define a complex valued potential as
\begin{equation}
    V(r) = V_0 \frac{\delta(r-r_0)}{r_0} =  P_0 \frac{\delta(r-r_0)}{r_0}  \left[  \frac{2G + i R}{2\Gamma} +  \frac{G}{W} \right].
\end{equation}
For brevity, we will define $V_0 = (\kappa+iR)P_0/(2\Gamma)$ with the parameter $\kappa = 2G(1+\Gamma/W)$ generalizing the Hermitian effect of non-resonant pumping on the condensate.
The solution for the 2D non-Hermitian Schr\"{o}dinger equation
\begin{equation} \label{gpe}
    i\hbar {\partial\Psi\over\partial t}  = \left[ - { \hbar^2 \nabla^2 \over 2 m^*} + V(r) -  {i \hbar \gamma \over 2}  \right] \Psi,
\end{equation}
in such a potential has the form
\begin{equation}
    \Psi(r,\varphi,t)=Ne^{il\varphi-iEt/\hbar}\times\left\lbrace \begin{matrix}
        cJ_l(kr), & k<r_0 \\
        H_l^{(1)}(kr), & k>r_0
    \end{matrix}\right.,
\end{equation}
where $k = \sqrt{2m^*(E+i\hbar \gamma/2)}/\hbar$ is the characteristic complex wave-vector simultaneously accounting for both radial expansion and decay, $c=H_l^{(1)}(kr_0)/J_l(kr_0)$ is the coefficient ensuring wavefunction continuity at $r=r_0$, and $N$ is the normalization factor.
The latter may be expressed as
\begin{equation}
    N = \sqrt{m^* \gamma \over 2 \pi \hbar} \left[
    \int_0^{\rho_0}  |c J_l(\rho\sqrt{\varepsilon+i})|^2 \rho d\rho  +  \int_{\rho_0}^\infty |H_l^{(1)}(\rho\sqrt{\varepsilon+i})|^2 \rho d\rho
    \right]^{-{1/2}}
\end{equation}
in normalized dimensionless radial coordinate $\rho = r \sqrt{m^*\gamma / \hbar}$, trap radius $\rho_0 = r_0 \sqrt{m^*\gamma/\hbar}$, and energy
$\varepsilon = {2 E / (\hbar \gamma)}$, as well as the factor $c= H_l^{(1)}(\rho_0\sqrt{\varepsilon+i})/J_l(\rho_0\sqrt{\varepsilon+i})$.

\begin{figure}
    \centering
    \includegraphics[width=0.8\linewidth]{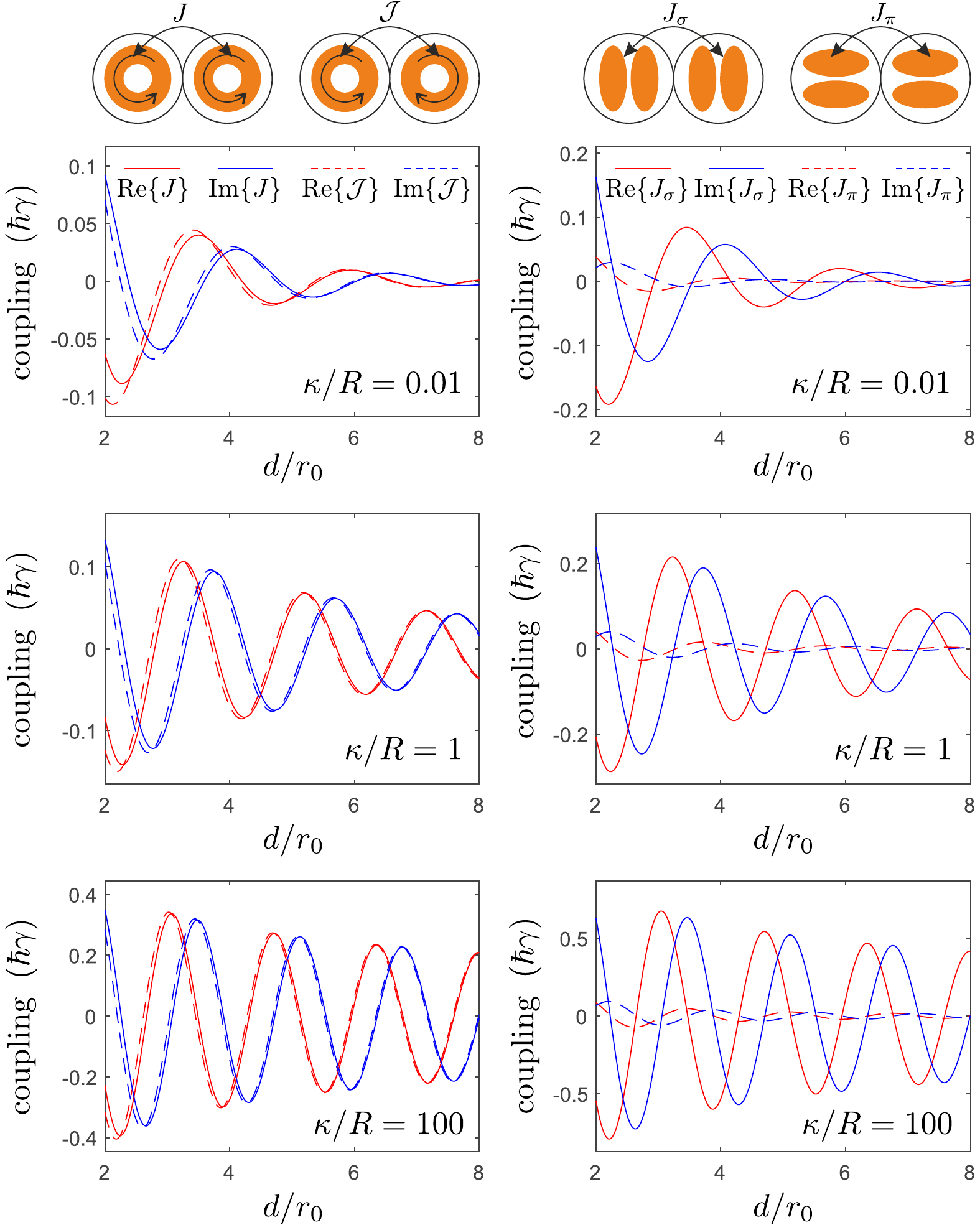}
    \caption{The coupling parameters $J$, $\mathcal{J}$ (left column) and $J_\sigma$, $J_\pi$ (right column) in units of $\hbar\gamma$ as a function of distance between trap $d$ in units of the trap radius $r_0$ at the threshold pumping power.
    The three rows correspond to different values of Hermitian to anti-Hermitian interaction parameter ratio $\kappa/R$.
    For each value the trap radius is chosen at the centre of the range, where the condensation occurs at the vortex state $l=\pm1$.
    Top row: $\kappa/R=0.01$, $\rho_0 = 1.57$, $\varepsilon = 2.61$, $p = 3.77$.
    Middle row: $\kappa/R=1$, $\rho_0 = 0.85$, $\varepsilon = 14.3$, $p = 4.37$.
    Bottom row: $\kappa/R=100$, $\rho_0 = 0.01$, $\varepsilon = 98026$, $p = 5.02$.}.
    \label{fig1}
\end{figure}

Introducing dimensionless pumping power $p = m^* P_0 R / (\hbar^2 \Gamma)$ one may also express the complex potential strength as
\begin{equation}
    V_0 = {\hbar^2 \over 2 m^*} p \left( {\kappa \over R} + i \right).
\end{equation}
Considering two adjacent identical traps $V_1(\mathbf{r}) = V(r_1)$ and $V_2(\mathbf{r}) = V(r_2)$, where the condensate is formed at $l=\pm1$, we assume the synchronized condensate wavefunction of the form
\begin{align}
    \Psi(\mathbf{r},t) & = \Psi_1(\mathbf{r},t) + \Psi_2(\mathbf{r},t) \\
    &=\left( \Psi_0(r_1) \left[ \psi_{1,+}  e^{+i\varphi_1} + \psi_{1,-} e^{-i\varphi_1} \right]+\Psi_0(r_2) \left[ \psi_{2,+} e^{+i\varphi_2} + \psi_{2,-} e^{-i\varphi_2} \right] \right) e^{-iEt/\hbar}.
\end{align}
Setting the origin of coordinates at the centre of the first trap we have
\begin{align}
    r_1 = x^2 + y^2, & \qquad \varphi_1 = \arg (x+iy) \\
    r_2 = (x-d)^2+y^2, & \qquad \varphi_2 = \arg (x-d+iy).
\end{align}
The two basis functions $\Psi_1$ and $\Psi_2$ satisfy
\begin{equation}
    [\hat{T}+V_{1,2}(\mathbf{r})-i\hbar\gamma/2] \Psi_{1,2} = E \Psi_{1,2}
\end{equation}
For the double trap Hamiltonian we have
\begin{align}
    [\hat{T}+V_1(\mathbf{r})+V_2(\mathbf{r})-i\hbar\gamma/2][\Psi_1(\mathbf{r})+\Psi_2(\mathbf{r})] = \\
    E[\Psi_1(\mathbf{r})+\Psi_2(\mathbf{r})] + V_1(\mathbf{r})\Psi_2(\mathbf{r}) + V_2(\mathbf{r})\Psi_1(\mathbf{r}).
\end{align}
Therefore for the time-dependent equation we have
\begin{align}
    i\hbar {d\over dt} & \left[ (\psi_{1,+} e^{+i\varphi_1}
   + {\psi_{1,-}} e^{-i\varphi_1})\Psi_0(r_1)  + ({\psi_{2,+}} e^{+i\varphi_2} + {\psi_{2,-}} e^{-i\varphi_2})\Psi_0(r_2) \right] = \\
    & V(r_1) ({\psi_{2,+}} e^{+i\varphi_2} + {\psi_{2,-}} e^{-i\varphi_2})\Psi_0(r_2)
    +V(r_2) ({\psi_{1,+}} e^{+i\varphi_1} + {\psi_{1,-}} e^{-i\varphi_1})\Psi_0(r_1).
\end{align}
Consequently integrating the equation above multiplied by $\Psi_{1,2}^*e^{\pm i \varphi}$ we can write down the equation on the vector $\psi=\psi_{1,+},\psi_{1,-},\psi_{2,+},\psi_{2,-}]^T$:
\begin{equation}
    i\hbar \partial_t \psi = -\left[ \begin{matrix}
    0 & 0 & J & \mathcal{J} \\
    0 & 0 & \mathcal{J} & J \\
    J & \mathcal{J} & 0 & 0 \\
    \mathcal{J} & J & 0 & 0 \\
    \end{matrix} \right] \psi,
\end{equation}
with the coupling parameters $J$ and $\mathcal{J}$ defined as
\begin{equation}
    J = -{\hbar^2 \over 2 m^*} N^2 p \left( { \frac{\kappa}{R}} + i \right) \left[ H_1^{(1)}(\rho_0\sqrt{\varepsilon+i})\right]^* \int_0^{2\pi} e^{i[\varphi_2(\varphi)-\varphi]}H_1^{(1)}\left(\rho_0\sqrt{\varepsilon+i}r_2(\varphi)/r_0\right)d\varphi,
\end{equation}
\begin{equation}
    \mathcal{J} = -{\hbar^2 \over 2 m^*} N^2 p \left( {\kappa \over R} + i \right) \left[ H_1^{(1)}(\rho_0\sqrt{\varepsilon+i})\right]^* \int_0^{2\pi} e^{i[\varphi_2(\varphi)+\varphi]}H_1^{(1)}\left(\rho_0\sqrt{\varepsilon+i}r_2(\varphi)/r_0\right)d\varphi,
\end{equation}
where
\begin{align}
    r_2(\varphi) = \sqrt{r_0^2 \sin^2\varphi + (d-r_0\cos\varphi)^2}, \;
    \varphi_2(\varphi) = \arg(r_0\cos\varphi - d + ir_0\sin\varphi).
\end{align}
In the basis of dipole states $\psi=(1/\sqrt{2})[\psi_{1,+}+\psi_{1,+},\psi_{1,+}-\psi_{1,+},\psi_{2,+}+\psi_{2,+},\psi_{2,-}-\psi_{2,+}]^T$, a similar linear equation reads:
\begin{equation}
    i\hbar \partial_t \psi^\prime = -\left[ \begin{matrix}
    0 & 0 & J_\sigma & 0 \\
    0 & 0 & 0 & J_\pi \\
    J_\sigma & 0 & 0 & 0 \\
    0 & J_\pi & 0 & 0 \\
    \end{matrix} \right] \psi^\prime,
\end{equation}
where the two coupling parameters $J_\sigma = J + \mathcal{J}$, $J_\pi = J - \mathcal{J}$ characterize the bonding strength of $\sigma$ and $\pi$ oriented dipole couples (see Fig. \ref{fig1}).

In the basis of vortices, arbitrary orientation of the two traps, described with the angle $\Theta$, results in the symmetry allowed rotation factors to the vorticity switching matrix elements:
\begin{equation}
    i\hbar \partial_t \psi = -\left[ \begin{matrix}
    0 & 0 & J & \mathcal{J}e^{-2i\Theta} \\
    0 & 0 & \mathcal{J}e^{2i\Theta} & J \\
    J & \mathcal{J}e^{2i\Theta} & 0 & 0 \\
    \mathcal{J}e^{-2i\Theta} & J & 0 & 0 \\
    \end{matrix} \right] \psi,
\end{equation}
In both bases the real and imaginary parts of the coupling parameters correspond to Hermitian and anti-Hermitian types of interaction.

Computed values of all coupling parameters are shown in Fig. \ref{fig1}.
We addressed the limiting cases of predominantly dissipative ($\kappa/R\ll1$) and tunneling ($\kappa/R\gg1$) types of interaction between the condensate and the reservoir, as well as the intermediate case where the two contributions are comparable.
In all three scenarios, the trap radius $r_0$ was chosen at the centre of the interval of values, corresponding to polariton condensation in the vortex states $l=\pm1$.
The corresponding dimensionless pumping power values $p$ and condensate energies $\varepsilon$ were numerically computed using the model in Ref. \cite{Cherotchenko2021}.
All coupling parameters exhibit the dependence on the distance $d$ between the traps of the type of damped oscillations.
In the vortex coupling representation, both parameters $J$ and $\mathcal{J}$ follow the same trend, while in the picture of coupled dipoles, $\sigma$ bonding strength is generally dominant over $\pi$ bonding.
Interestingly, the Josephson-type coupling, governed by particle exchange between the two condensates, and the dissipative mechanism contribution to coupling, quantified with the real and imaginary parts of the coupling parameters respectively, are varying with the phase shift close to $\pi/4$ with the distance $d$.

\section{Polariton vortex Tight Binding Hamiltonian}
\subsection{The infinite honeycomb lattice (spinor polaritons)}\label{subsec:infinite_spinor}
We can write a tight binding Hamiltonian to describe the nearest neighbour coupling of localized polariton vortex states arranged in a honeycomb lattice~\cite{castro_neto_electronic_2009,leggett_lecture_2010,utermohlen_tight-binding_2018}. Each lattice site of $A$ and $B$ is home to a vortex and antivortex mode, labelled by $+$ and $-$ respectively. As each vortex and antivortex is coupled to both its co-rotating and counter-rotating neighbours with energies $J$ and $\mathcal{J}$ respectively, the tight binding Hamiltonian follows:
\begin{equation}
\begin{split}
     \hat{H} = -\sum_{i \in A} \sum_{\delta} \biggl[\underbrace{J\hat{a}^\dagger_{i,+}\hat{b}_{i + \delta, +}}_\text{vortex to vortex}+~ \underbrace{\mathcal{J}\hat{a}^\dagger_{i,+}\hat{b}_{i + \delta, -}e^{-2i\Theta_{i,i+\delta}}}_\text{vortex to antivortex} \\
     +~ \underbrace{J\hat{a}^\dagger_{i,-}\hat{b}_{i + \delta, -}}_\text{antivortex to antivortex}+~ \underbrace{\mathcal{J}\hat{a}^\dagger_{i,-}\hat{b}_{i + \delta, +}e^{2i\Theta_{i,i+\delta}}}_\text{antivortex to vortex}+~\text{h.c.} \biggr]
\end{split}
\end{equation}

\noindent
where $i$ and $(i+\delta)$ label the lattice sites in the sublattice $A(B)$, $\Theta$ is the angle between the connecting centres of the two nearest neighbours to the positive $x$ axis. Explicitly, $\Theta_{i,i+\delta_1} = \pi/3$, $\Theta_{i,i+\delta_2} = -\pi/3$ and $\Theta_{i,i+\delta_3} = \pi$. 
$\hat{a}_{i,\pm}^{(\dagger)}$ and $\hat{b}_{i+\delta,\pm}^{(\dagger)}$ represent the annihilation (creation) operators of a particle in the vortex or antivortex mode (given by $\pm$) at lattice sites $i$ and $i + \delta$ respectively. Here, $\text{h.c.}$ stands for \textit{Hermitian conjugate}.

The creation operator in momentum space is defined as:
\begin{equation}
    a^{\dagger}_{i,\pm} = \frac{1}{\sqrt{N_l/2}} \sum_{\boldsymbol{k}} e^{i\boldsymbol{k}\cdot\boldsymbol{r}_i}\hat{a}^{\dagger}_{\boldsymbol{k},\pm},
    \label{eq:FT_chain}
\end{equation}
where $N_l$ is the number of lattice sites. This allows us to re-write the Hamiltonian as:
\begin{align}
\begin{split}
    \hat{H} = -\frac{1}{N_l/2} \sum_{i\in A} \sum_{\boldsymbol{\delta},\boldsymbol{k},\boldsymbol{k}^\prime} \Bigg[&e^{i\boldsymbol{k}\cdot\boldsymbol{r}_i}e^{-i\boldsymbol{k}^\prime \cdot(\boldsymbol{r}_i + \boldsymbol{\delta})} \bigg(J\hat{a}^\dagger_{\boldsymbol{k},n+}\hat{b}_{\boldsymbol{k}^\prime,n+} + \mathcal{J}\hat{a}^\dagger_{\boldsymbol{k},n+}\hat{b}_{\boldsymbol{k}^\prime,n-}e^{-2i\Theta_{\boldsymbol{\delta}}} \\&~~~+ J\hat{a}^\dagger_{\boldsymbol{k},n-}\hat{b}_{\boldsymbol{k}^\prime,n-} + \mathcal{J}\hat{a}^\dagger_{\boldsymbol{k},n-}\hat{b}_{\boldsymbol{k}^\prime,n+}e^{2i\Theta_{\boldsymbol{\delta}}}\bigg) ~ +~ \text{h.c}\Bigg]
    \end{split}
    \end{align}
    
    \begin{align}
    \begin{split}
         = - \sum_{\boldsymbol{\delta},\boldsymbol{k}}  &\Bigg[e^{-i\boldsymbol{k}\cdot\boldsymbol{\delta}} 
    \bigg(J\hat{a}^\dagger_{\boldsymbol{k},+}\hat{b}_{\boldsymbol{k},+} + \mathcal{J}\hat{a}^\dagger_{\boldsymbol{k},+}\hat{b}_{\boldsymbol{k},-}e^{-2i\Theta_{\boldsymbol{\delta}}} + J\hat{a}^\dagger_{\boldsymbol{k},-}\hat{b}_{\boldsymbol{k},-} + \mathcal{J}\hat{a}^\dagger_{\boldsymbol{k},-}\hat{b}_{\boldsymbol{k},+}e^{2i\Theta_{\boldsymbol{\delta}}}\bigg)\\
    + &~~~e^{i\boldsymbol{k}\cdot\boldsymbol{\delta}}\bigg(J\hat{b}^\dagger_{\boldsymbol{k},+}\hat{a}_{\boldsymbol{k},+} + \mathcal{J}\hat{b}^\dagger_{\boldsymbol{k},-}\hat{a}_{\boldsymbol{k},+}e^{2i\Theta_{\boldsymbol{\delta}}} + J\hat{b}^\dagger_{\boldsymbol{k},-}\hat{a}_{\boldsymbol{k},-} + \mathcal{J}\hat{b}^\dagger_{\boldsymbol{k},+}\hat{a}_{\boldsymbol{k},-}e^{-2i\Theta_{\boldsymbol{\delta}}}\bigg)\Bigg],
    \end{split}
\end{align}
where we have used $\sum_{i\in A}e^{i(\boldsymbol{k} - \boldsymbol{k}^\prime)\cdot\boldsymbol{r}_i} = \frac{N_l}{2}\delta_{\boldsymbol{kk}^\prime}$ to simplify the Hamiltonian. 

Next, we include the expressions:
\begin{equation}
    \Delta_{\boldsymbol{k}} = \sum_{\boldsymbol{\delta}} e^{-i\boldsymbol{k}\cdot\boldsymbol{\delta}}
\end{equation}
\begin{equation}
    \Delta_{\boldsymbol{k}\pm} = \sum_{\boldsymbol{\delta}} e^{-i\boldsymbol{k}\cdot\boldsymbol{\delta}\mp2i\Theta_{i,i+\delta}}
\end{equation}
allowing us to express the tight-binding Hamiltonian as:
\begin{equation}
    \hat{H} = \sum_{\boldsymbol{k}} \Psi^\dagger \boldsymbol{h}(\boldsymbol{k}) \Psi
\end{equation}
where $\boldsymbol{\Psi}^\dagger = (\hat{a}^\dagger_{\boldsymbol{k},+} , \hat{a}^\dagger_{\boldsymbol{k},-} , \hat{b}^\dagger_{\boldsymbol{k},+} , \hat{b}^\dagger_{\boldsymbol{k},-})$ and,
\begin{equation}
    \boldsymbol{h}(\boldsymbol{k}) = \begin{pmatrix}
    0&0& -J\Delta_{\boldsymbol{k}}&-\mathcal{J}\Delta_{\boldsymbol{k}+}\\
    0&0& -\mathcal{J}\Delta_{\boldsymbol{k}-}&-J\Delta_{\boldsymbol{k}}\\
    -J\Delta^*_{\boldsymbol{k}}&-\mathcal{J}\Delta_{\boldsymbol{k}-}^* &0&0\\
    -\mathcal{J}\Delta^*_{\boldsymbol{k}+}&-J\Delta^*_{\boldsymbol{k}} &0&0
    \end{pmatrix}.
    \label{eq:Hvortex}
\end{equation}
The top-right block of the above operator corresponds to equation (13) in the main text.

\subsection{The finite honeycomb strip}\label{subsec:finite_spinor}
We consider the honeycomb lattice of coupled vortex modes, where the lattice is infinite in the $y$ direction, but finite in $x$. We decompose the lattice into $N$ infinite length zigzag chains that consist of alternating $A,B$ sublattice sites separated by alternating $\boldsymbol{\delta}_{1},-\boldsymbol{\delta}_{2}$ translations, as shown in the red box of Figure~1(a) in the main text. The neighbouring chains are then connected from $A$ sites of one chain to the $B$ sites of a neighbouring chain translated by $-\boldsymbol{\delta}_3$. 

The Hamiltonian of each zigzag chain follows \autoref{eq:Hvortex}, but with 
\begin{align}
    \Delta_{\boldsymbol{k}} = e^{-i\boldsymbol{k}\cdot\boldsymbol{\delta}_1} + e^{-i\boldsymbol{k}\cdot\boldsymbol{\delta}_2} = 2e^{-ik_x\frac{a}{2}}\cos{\left(k_y \frac{\sqrt{3}a}{2}\right)},
    \label{eq:dk_y}
\end{align}
and
\begin{align}
     \Delta_{\boldsymbol{k}\pm} =   e^{-i\boldsymbol{k}\cdot\boldsymbol{\delta}_1}e^{\mp 2i\Theta_{\boldsymbol{\delta}_1}} + e^{-i\boldsymbol{k}\cdot\boldsymbol{\delta}_2}e^{\mp 2i\Theta_{\boldsymbol{\delta}_2}} = 2e^{-ik_x\frac{a}{2}}\cos{\left(k_y \frac{\sqrt{3}a}{2} \pm \frac{2\pi}{3}\right)}.\label{dkpm_y}
\end{align}
To differentiate from the Hamiltonian for the infinite lattice, we label the Hamiltonian of a single infinite zigzag chain as $\boldsymbol{h}_{\textrm{ch}}$. Each $B$ sublattice site in chain $n$ is coupled to the $A$ lattice sites of chain $n+1$ which are translated by $-\boldsymbol{\delta}_3$. Note that if the labeling of the chains was reversed (i.e. increasing from right to left), the translation would be be $\boldsymbol{\delta}_3$ from $A$ sites to $B$ sites. This coupling follows the matrix:
\begin{equation}
    \boldsymbol{T}(\boldsymbol{k}) = \begin{pmatrix}
    0&0&0&0\\
    0&0&0&0\\
    -Je^{-ik_xa}&-\mathcal{J}e^{-ik_xa}&0&0\\
    -\mathcal{J}e^{-ik_xa}&-Je^{-ik_xa}&0&0
    \end{pmatrix}.
\end{equation}
We can now construct a $4N \times 4N$ matrix Hamiltonian, $\boldsymbol{H}_\textrm{chain}(\boldsymbol{k})$, to describe the whole system of $N$ coupled chains, given by:
\begin{equation}
    \hat{H}(\boldsymbol{k}) = \sum_{\boldsymbol{k}}\boldsymbol{\Psi}^\dagger \boldsymbol{H}_\textrm{chain}(\boldsymbol{k}) \boldsymbol{\Psi}
\end{equation}
where $\boldsymbol{\Psi}^\dagger = (\boldsymbol{\Psi}^\dagger_1 ,  \boldsymbol{\Psi}^\dagger_2 , \cdots , \boldsymbol{\Psi}^\dagger_{4N})$. We also introduce the $4\times4$ zero matrix as $\boldsymbol{0} = 0_{4,4}$. Each of the $N$ row-blocks of $\boldsymbol{H}_\textrm{chain}(\boldsymbol{k})$ indexed by $r$ are then given by:
\begin{align}
    r=1: & ~~~\Big(\boldsymbol{h}_{\textrm{ch}}~~\boldsymbol{T}~~ \boldsymbol{0}_{\times(N-2)}\Big)\label{eq:H_1}\\
    1<r<N: & ~~~\Big(\boldsymbol{0}_{\times(r-2)}~~\boldsymbol{T}^\dagger~~\boldsymbol{h}_{\textrm{ch}}~~\boldsymbol{T}~~ \boldsymbol{0}_{\times(N - r -1)}\Big)\label{eq:H_2}\\
    r = N: & ~~~\Big(\boldsymbol{0}_{\times(N-2)}~~\boldsymbol{T}^\dagger~~ \boldsymbol{h}_{\textrm{ch}}\Big)\label{eq:H_3}
\end{align}
\noindent
where the subscript $_{\times(X)}$ means to $\boldsymbol{0}$ appears $X$ times adjacently in that block of row $r$. For example, if we choose $N = 3$, then we can write $\boldsymbol{H}_\textrm{chain}(\boldsymbol{k})$ as:

\begin{gather}
    \boldsymbol{H}_\textrm{chain}(\boldsymbol{k}) = \begin{pmatrix}
    \boldsymbol{h}_\textrm{ch} & \boldsymbol{T} & \boldsymbol{0}\\
    \boldsymbol{T}^\dagger & \boldsymbol{h}_\textrm{ch} & \boldsymbol{T} \\
    \boldsymbol{0} & \boldsymbol{T}^\dagger & \boldsymbol{h}_\textrm{ch}
    \end{pmatrix}.
    \label{eq:H_manyN}
\end{gather}

\subsection{The finite honeycomb strip (bogolon excitation)}\label{subsec:finite_bog}
Constructing the bogolon operator for a finite strip of polariton vortex honeycomb follows the same principles as in section 2.3.1 in the main text. From Eq.~12 in the main text we write the coupling between sublattice sites in an infinite zigzag chain as,
\begin{equation}
\mathbf{J}_{\boldsymbol{k},\textrm{ch}} = (\hat{J}_{1} e^{-ik_1} + \hat{J}_{2} e^{-ik_2}) \otimes \hat{\sigma}_z
\label{eq:Jkri_finite}
\end{equation}
such that the evolution of bogolons along each chain is described by,
\begin{equation} 
\mathbf{M}_{\textrm{ch}} = \begin{pmatrix} \mathcal{M}_A & \mathbf{J}_{\boldsymbol{k},\textrm{ch}}  \\
\mathbf{J}_{\boldsymbol{k},\textrm{ch}}^\dagger   &  \mathcal{M}_B 
   \end{pmatrix}.
   \label{eq:Mfinite}
\end{equation}
The coupling between each chain is then given by,
\begin{equation} 
\mathbf{M}_{\textrm{T}} = \begin{pmatrix} 
0 & 0  \\
\mathbf{J}_{\boldsymbol{k},\textrm{T}}^\dagger   &  0 
   \end{pmatrix},
   \label{eq:MT}
\end{equation}
where 
\begin{equation}
\mathbf{J}_{\boldsymbol{k},\textrm{T}} = \hat{J}_{0} \otimes \hat{\sigma}_z.
\label{eq:Jkri_finite}
\end{equation}
We then construct an $8N \times 8N$ matrix Hamiltonian, $\boldsymbol{H}_\textrm{bog}(\boldsymbol{k})$, following:
\begin{equation}
    \hat{H}_{\textrm{bog}}(\boldsymbol{k}) = \sum_{\boldsymbol{k}}\boldsymbol{\Psi}^\dagger \boldsymbol{H}_\textrm{bog}(\boldsymbol{k}) \boldsymbol{\Psi}
\end{equation}
where
\begin{equation}
    \boldsymbol{\Psi} = \begin{pmatrix}
        \boldsymbol{\Psi}_1\\
        \boldsymbol{\Psi}_2\\
        \vdots\\
        \boldsymbol{\Psi}_{8N}\\
    \end{pmatrix},
    \qquad 
    \boldsymbol{\Psi}^\dagger = \begin{pmatrix}
    \boldsymbol{\Psi}^\dagger_1 &  \boldsymbol{\Psi}^\dagger_2 & \cdots & \boldsymbol{\Psi}^\dagger_{8N}
    \end{pmatrix}.
\end{equation}
Each of the $N$ rows of $\boldsymbol{H}_\textrm{bog}(\boldsymbol{k})$, consisting of $N$ $8\times 8$ matrix blocks and indexed by $r$, are given by:
\begin{align}
    r=1: & ~~~\Big(\mathbf{M}_{\textrm{ch}}~~\mathbf{M}_{\textrm{T}}~~ \boldsymbol{0}_{\times(N-2)}\Big)\label{eq:H_1}\\
    1<r<N: & ~~~\Big(\boldsymbol{0}_{\times(r-2)}~~\mathbf{M}_{\textrm{T}}^\dagger~~\mathbf{M}_{\textrm{ch}}~~\mathbf{M}_{\textrm{T}}~~ \boldsymbol{0}_{\times(N - r -1)}\Big)\label{eq:H_2}\\
    r = N: & ~~~\Big(\boldsymbol{0}_{\times(N-2)}~~\mathbf{M}_{\textrm{T}}^\dagger~~ \mathbf{M}_{\textrm{ch}}\Big)\label{eq:H_3}
\end{align}
\noindent
where $\boldsymbol{0} = 0_{8,8}$. For example, if we choose $N = 3$, then $\boldsymbol{H}_{\textrm{bog}}(\boldsymbol{k})$ is defined as:
\begin{gather}
    \boldsymbol{H}_{\textrm{bog}}(\boldsymbol{k}) = \begin{pmatrix}
    \boldsymbol{M}_\textrm{ch} & \boldsymbol{M}_\textrm{T} & \boldsymbol{0}\\
    \boldsymbol{M}_{\textrm{T}}^\dagger & \boldsymbol{M}_\textrm{ch} & \boldsymbol{M}_\textrm{T} \\
    \boldsymbol{0} & \boldsymbol{M}_{\textrm{T}}^\dagger & \boldsymbol{M}_\textrm{ch}
    \end{pmatrix}.
    \label{eq:H_manyNBog}
\end{gather}


\section{Comparison of vortex bogolon edge modes with spinor polaritons}
In this section, we will show that the dispersion of the vortex bogolons can, for certain ranges of parameters, quantitatively match the dispersion of the conventional spinor polariton Chern insolator in photonic graphene~\cite{nalitov_polariton_2015}. The main differences are: First, the state space dimension for bogolons is doubled compared to spinor polaritons. This is because both positive and negative energy bogolons must be kept track of in the linearlization procedure. Second, instead of a real magnetic field which splits the polariton spin states (through the exciton wavefunction), here the splitting between $l=\pm 1$ bogolons stems from the underlying chiral FM vortex lattice. For the latter, the strength of the splitting is proportional to the condensate amplitude (or pump power). We show that the spinor polariton graphene spectra for a range of applied magnetic fields can be qualitatively reproduced by the bogolon system using a corresponding range of vortex amplitudes. 
\begin{figure}
    \centering
\includegraphics[width=0.99\textwidth]{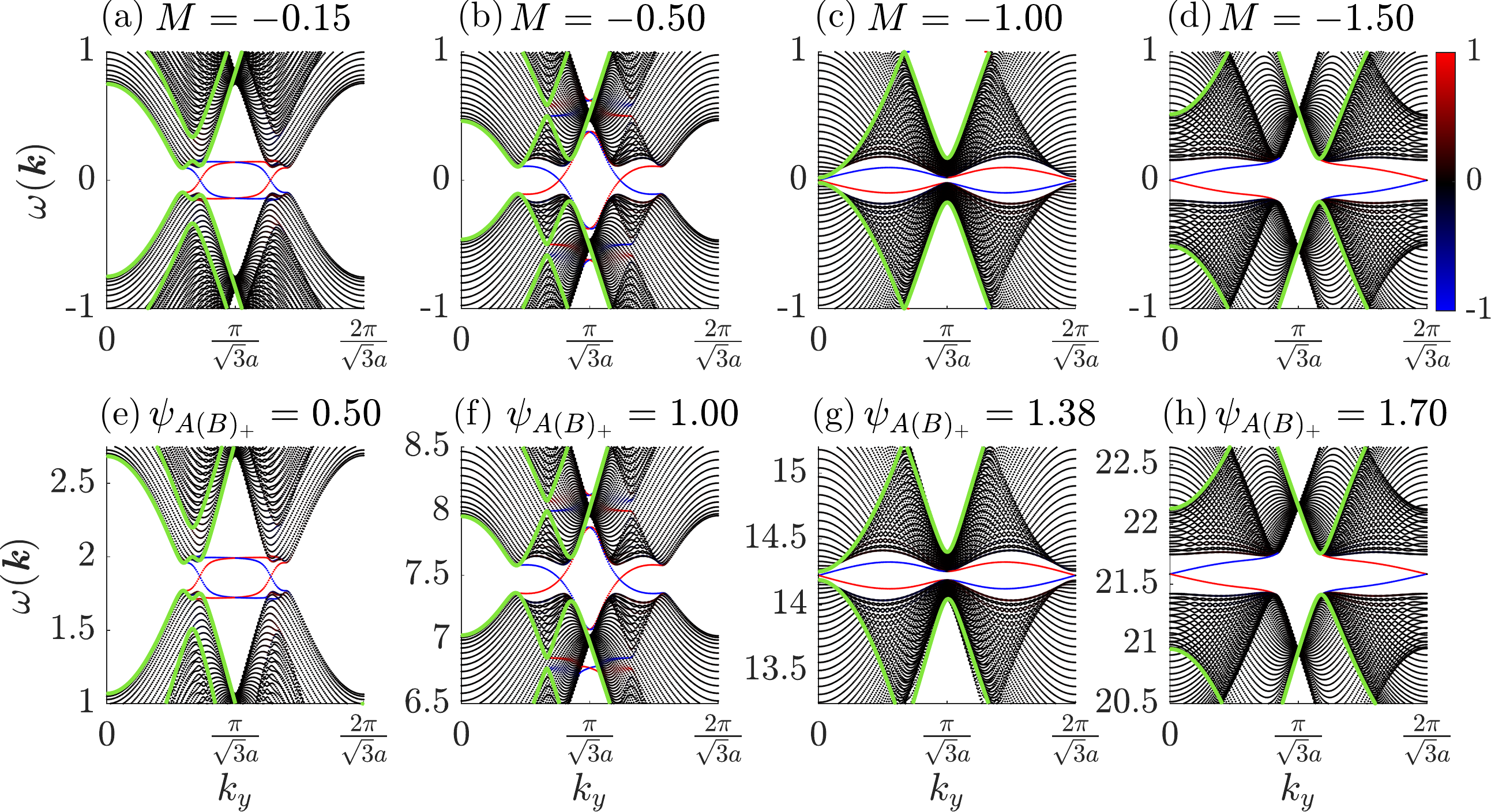}
    \caption{Dispersion comparison between (a-d) the spinor tight binding model used in \cite{nalitov_polariton_2015} and the (e-h) vortex bogolon operator developed in this work for both the (green) infinite and (blue/black/red) $N=50$ finite honeycomb lattices. Here, the colour of the finite spectra corresponds to the localisation of the bands to the (blue) left and (red) right lattice edges. In all plots, $k_x = 2\pi/3a$, $J = 1$ and $\mathcal{J} = 0.1$. In panels (e-h) we have set $\alpha = 4$, $\psi_{A(B),-}=0$ and, respectively, $p = [0.25, \ 1, \ 1.9, \ 2.9]$.}
    \label{fig:TB_Bog}
\end{figure}

As we mentioned in the main text. The coupling between vortices is analogous to spinor polaritons in a micropillar lattice with TE-TM splitting~\cite{nalitov_polariton_2015}. This means that spinor polaritons can also be described by~\eqref{eq:Hvortex} and the influence of the out-of-plane magnetic field can be simply appended to the right hand site as an operator $\dots + M \hat{\sigma}_0 \otimes \hat{\sigma}_z$ where $M$ is the field strength. In \autoref{fig:TB_Bog}(a-d), the energy spectra from the spinor system in Ref.~\cite{nalitov_polariton_2015} are shown over a range of external magnetic field strengths for the (green) infinite and (blue/black/red) finite honeycomb lattices. In all plots, a gap has opened between the Dirac points. We see that in the finite lattice, bands cross between the conduction and valance bands of the bulk, corresponding to the edge states, as indicated by their colour. Interestingly, we are able to qualitatively reproduce all of these energy spectra using the bogolon operator, where instead of varying an external magnetic field strength, the induced magnetic field is controlled by the amplitude of the vortex lattice, $\psi_{A(B),+}$, as shown in \autoref{fig:TB_Bog}(e-h). Not only does this confirm that a co-rotating vortex lattice induces an effective magnetic field, but also that its strength is fully controlable via the vortex amplitudes. Experimentally, this is achievable through 
varying the intensity of the polariton condensate lattice through controlling the pump power of the ring-shaped optical traps.

\begin{figure}[h]
    \centering
\includegraphics[width=0.61\textwidth]{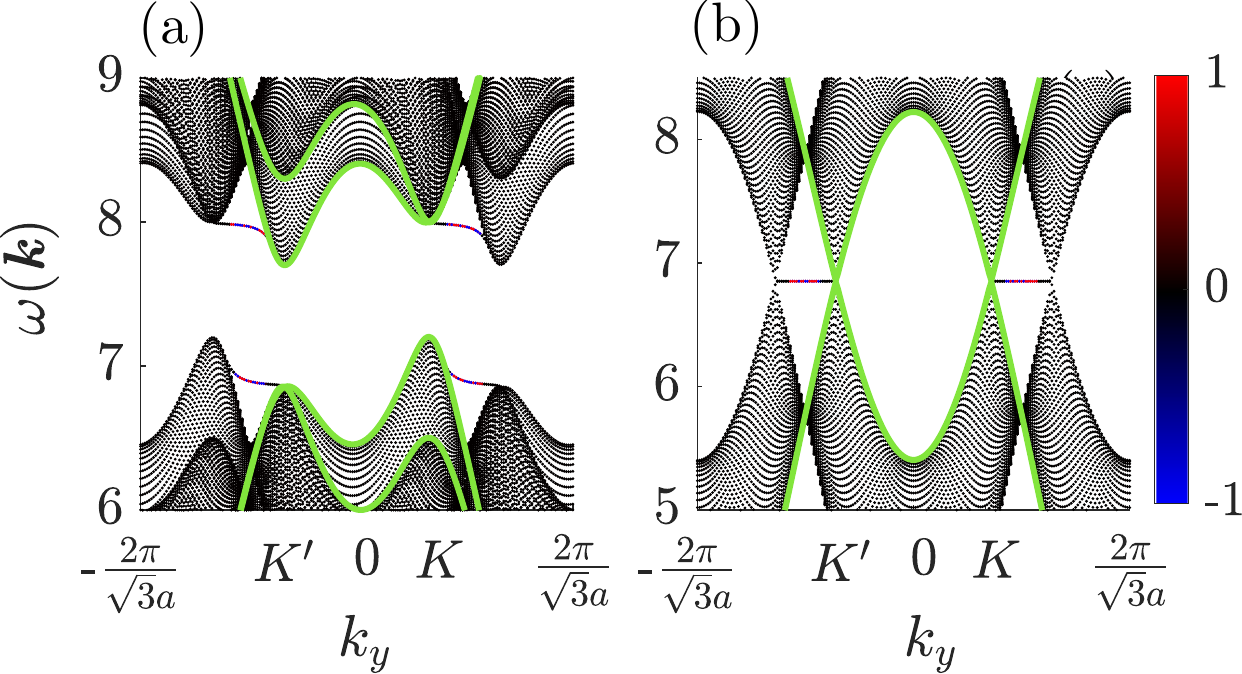}
    \caption{Real energy band structure of the (a) antiferromagnetic honeycomb vortex lattice  with $\psi_{A,+} = \psi_{B,-} = 1$, $\psi_{A,-} = \psi_{B,+} = 0$ and (b) dipole honeycomb vortex lattice  with $\psi_{A,\pm} = \psi_{B,\pm} = 1$ of the (blue/black/red) finite ($N = 50$) and (green) infinite Bogoliubov excitation Hamiltonian across a single Brillouin zone with $k_x = 2\pi/3a$. The colour of the finite lattice spectra corresponds to the localisation of the bands to the (blue) left and (red) right lattice edges. In both plots, $\alpha = 4$, $p=1$ and $\mathcal{J} = 0.1$.}
    \label{fig:AFM_dipole}
\end{figure}

\section{Other vortex arrangements}
Here, we check whether the topologically nontrivial bogolon gap opening, connected by chiral edge modes, is inherent to the chiral FM vortex solution discussed in the main text. To check this we try two other type of solutions we refer to as the AFM vortex solution ($\psi_{A,+} = \psi_{B,-} = 1, \psi_{A,-} = \psi_{B,+} = 0$) and the dipole solution ($\psi_{A,\pm} = \psi_{B,\pm} = 1$). In \autoref{fig:AFM_dipole} we repeat our calculations for bogolon dispersion in the honeycomb strip for (a) the AFM solution and (b) the dipole solution. For the former there is a bandgap but no edge states cross it implying that the system is topologically trivial. In the latter case there is no gap opening between the conduction and valence bands of the system. Thus, we conclude that indeed the co-rotating FM vortex lattice induces an internal magnetic field and is responsible for the gap opening in Fig.~2(b,d,f) shown in the main manuscript.

\section{2D GPE}
\begin{figure}[h]
    \centering    
    \includegraphics[width=0.65\textwidth]{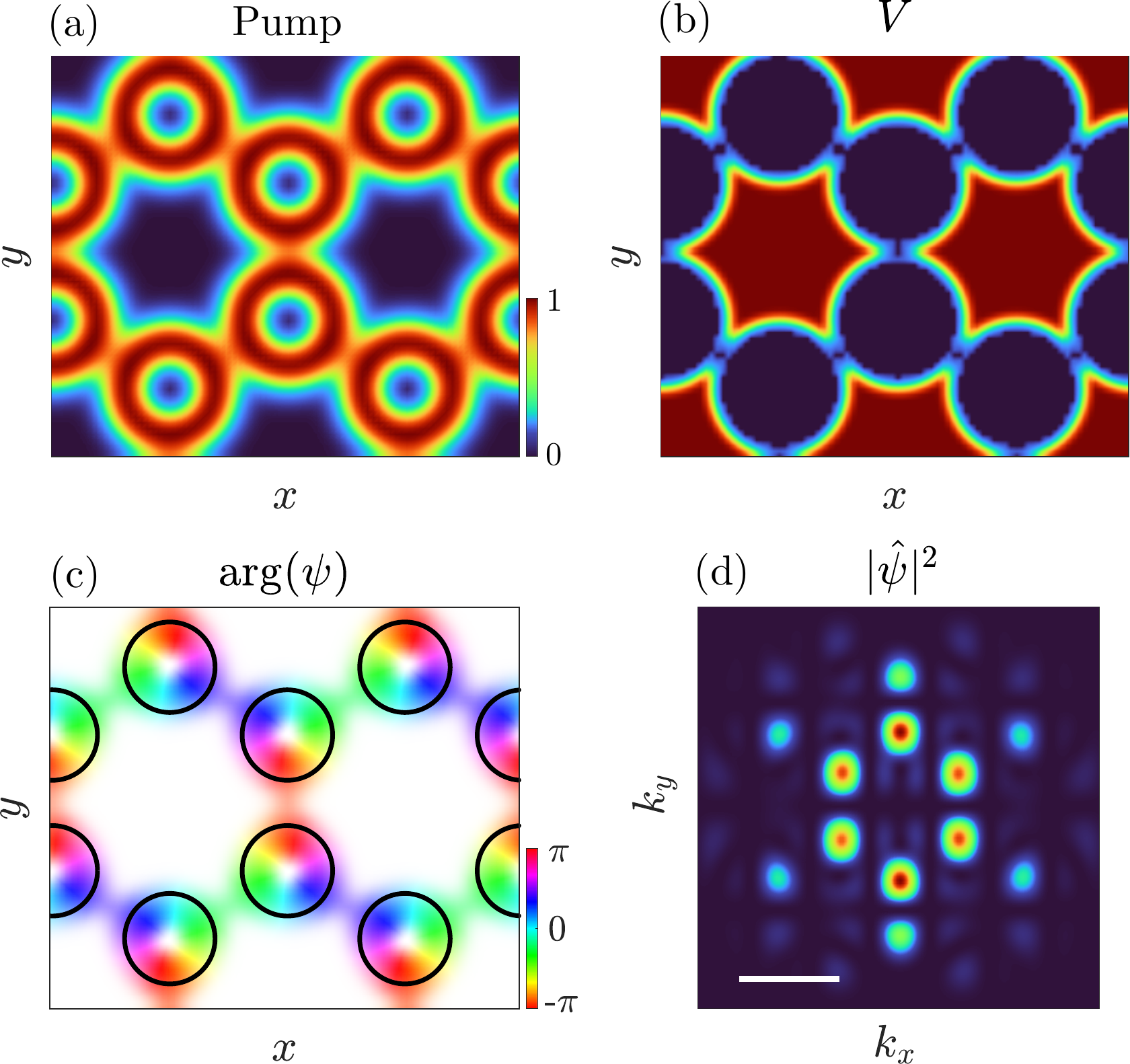}
    \caption{(a) Pump profile, (b) additional confinement potential $V$ and (c) steady state phase $\arg(\psi)$ of the 2D GPE honeycomb vortex lattice with a transparency map of $\vert \psi \vert^2$. The lattice is nonresonantly pumped and optically trapped by 8~$\upmu$m diameter rings with $a = 12$~$\upmu$m, indicated by black circles. (d) The corresponding normalised momentum space intensity of the wavefunction, $\vert \hat{\psi} \vert ^2$, where the white scalebar represents 0.5~$\upmu$m$^{-1}$.}
    \label{fig:vortexLattice}
\end{figure}

Using the 2D GPE as defined in \autoref{eq:GPE_psi}-\ref{eq:GPE_nx} we numerically demonstrate the feasibility of creating a honeycomb lattice of co-rotating polariton condensate vortices. Here, each vortex is nonresonantly excited by a 8~$\upmu$m diameter ring as shown in \autoref{fig:vortexLattice}(a), where this profile confines the first excited state condensate. By choosing an appropriate separation distance between nearest neighbours and an auxiliary confining potential $V(\mathbf{r})$ for stability reasons [see \autoref{fig:vortexLattice}(b)], we realise a lattice of co-rotating vortices, as shown in \autoref{fig:vortexLattice}(c) with a lattice constant of $a = 12$~$\upmu$m. In panel (c) we show the phase of the wavefunction with transparency proportional to its intensity. The corresponding momentum space intensity of the wavefunction is shown in \autoref{fig:vortexLattice}(d), where the three $\boldsymbol{K}$ and $\boldsymbol{K}^\prime$ are clearly seen in the hexagonal arrangement, characteristic of the traditional honeycomb lattice. Additionally, this plot shows a secondary outer hexagon of bright Dirac points, which we see in all energy spectra of the vortex honeycomb lattice and the tight binding spin-system with spin-orbit coupling. 

\end{document}